\documentclass[useAMS,usenatbib,usegraphicx]{mn2e}
\usepackage{journals}
\input{epsf}
\usepackage{color}
\usepackage{graphicx}
\usepackage[utf8]{inputenc}
\usepackage{caption}
\usepackage{subcaption}

\definecolor{darkgreen}{rgb}{0,0.7,0}
\usepackage[bottom]{footmisc}

\title
{A new method to measure the mass of galaxy clusters}

\author[Falco et al.]
{Martina  Falco$^1$,  Steen H. Hansen$^1$,
 Radoslaw Wojtak$^1$, Thejs~Brinckmann$^{1}$,
\newauthor
Mikkel~Lindholmer$^{1}$ and Stefania~Pandolfi$^{1}$\\
$^1$ Dark Cosmology Centre, Niels Bohr Institute, University of Copenhagen,
Juliane Maries Vej 30, 2100 Copenhagen, Denmark\\
{\rm falco@dark-cosmology.dk},\\
}

\begin{document}

\maketitle

\vspace{-1in}

\begin{abstract}

The mass measurement of galaxy clusters is an important tool for the
determination of
cosmological parameters describing the matter and energy content of
the Universe. However, the standard methods rely on various
assumptions about the shape or the level of equilibrium of the cluster.
We present a novel method of measuring cluster masses. It is
complementary to most of the other methods, since it only uses kinematical information
from outside the virialized cluster. 
Our method identifies 
objects, as galaxy sheets or filaments, in the cluster outer region, and infers the cluster mass by modeling how the massive cluster
perturbs the motion of the structures from the Hubble flow.
At the same time, this technique allows to constrain the three-dimensional orientation of the detected structures with a good accuracy.
We use a cosmological numerical simulation to test the method.
We then apply the method to the Coma cluster, where we find two
galaxy sheets, and measure the mass of Coma to be $M_{\rm vir}=(9.2\pm\,2.4)\times\,10^{14}\rm  M_\odot$,
in good agreement with previous measurements obtained with the
standard methods.

\end{abstract}


\vspace{-0.45in}

\section{Introduction}
The picture of the large-scale structures reveals that matter in the
Universe forms an intricate and complex system, defined as ``cosmic
web'' \citep{Z82,S83,E84,Bond96,a2010}.

First attempts of mapping the three-dimensional spatial distribution
of galaxies in the Universe \citep{g78,de86,geller89,Sh96}, as well as more recent large galaxy
surveys \citep{Col03,sdss,H05}, display a strongly anisotropic morphology. 
The galactic mass distribution seems to form a rich cosmos containing clumpy
structures, as clusters, sheets and filaments, surrounded by large voids \citep{vdw09}.  A similar cosmic network has emerged
from cosmological N-body simulations of the Dark Matter distribution \citep{Bond96,aragon2007,HAHN2007}.

The large scale structures are expected to span a range of scales that
goes from a few up to hundreds of megaparsec. Despite the many well-established methods to identify clusters and voids, there is not yet a
clear characterization of filaments and sheets. Due to their complex shape, there is not a common agreement on the definition and the internal properties of these objects \citep{B2010}. Moreover, their detection in observations is extremely difficult due to the projection
effects. Nevertheless, several automated algorithms for filament and sheet finding, both in 3D and 2D, have been developed \citep{N06,aragon2007,Sou08,B2010}. Several
galaxy filaments have been detected by eye \citep{C05,P08} and Dark
Matter filaments have also been detected from their weak gravitational
lensing signal \citep{D12}. Powerful methods for the cosmic web
classification, are based on the study of the Hessian of the
gravitational potential and the shear of the velocity field
\citep{HAHN2007,hoffman12}.

From the qualitative point of view, several elaborate theories have
been proposed. The cosmic web nature is intimately connected with the
gravitational formation process. In the standard model of hierarchical
structure formation, the cosmic structure has emerged from the growth
of small initial density fluctuations in the homogeneous early
Universe \citep{Peebles80,Davis85,WF91}. The accretion process
involves the matter flowing out of the voids, collapsing into sheets
and filaments, and merging into massive clusters. Thus, galaxy
clusters are located at the intersection of filaments and sheets, which operate as channels for the matter to flow into them \citep{van93,Colb99}. The innermost part of clusters tends to eventually reach the virial equilibrium.

As result of this gravitational collapse, clusters of galaxies are the
most recent structures in the Universe. For this reason, they are
possibly the most easy large-scale systems to study. Mass measurement
of galaxy clusters is of great interest for understanding the
large-scale physical processes and the evolution of structures in the
Universe \citep{W2010}. Moreover, the abundance of galaxy clusters as
function of their mass is crucial for constraining cosmological
models: the cluster mass function is an important tool for the
determination of the amount of Dark Matter in the Universe and for
studying the nature and evolution of Dark Energy
\citep{H01,C09,Allen11}.
The oldest method for the cluster mass determination is based on the
application of the virial theorem to positions and velocities of the
cluster members \citep{Z33}. 
This method suffers from the main limitation that the estimated mass
is significantly biased when the cluster is far from
virialization. More recent and sophisticated techniques also rely
strongly on the assumption of hydrostatic or dynamical equilibrium. 
The cluster mass profile can be estimated, for example, from
observations of density and temperature of the hot X-ray gas, through the application of the hydrostatic equilibrium equation \citep{Ettori02,Bor04,Zappa06,SA07,HH11}. Another
approach is based on the dynamical analysis of cluster-member galaxies
and involves the application of the Jeans equations for steady-state spherical system. \citep{Gir98,LM03,LW06,mamon10}. 
Additional cluster mass estimators have been proposed, which are
independent of the cluster dynamical state.
A measurement of the total cluster mass can be achieved by studying
the distortion of background galaxies due to gravitational lensing
\citep{M10,L11}. The lensing technique is very sensitive to the
instrument resolution and the projection effects.
The caustic method has been proposed by \cite{D99}. This method
requires very large galaxy surveys, in order to determine the caustic curve accurately.
Therefore, the development of new techniques and the combination of
different independent methods, is extremely useful for providing a more accurate cluster mass measurement.

The Coma cluster of galaxies (Abell 1656) is one of the most
extensively studied system of galaxies \citep{Biviano98}, as the most
regular, richest and best-observed in our neighborhood. The X-ray observations have provided several mass estimates
\citep{Hug89,Watt92}, obtained by assuming hydrostatic
equilibrium. Dynamical mass measurements with different methods, based
on the assumption of dynamical equilibrium, are reported in
\citep{The86,LM03}. \cite{Geller99} perform a dynamical measurement
of the Coma cluster, using the caustic method, and weak lensing mass estimates of Coma have been carried on by \cite{kubo07} and \cite{gavazzi09}.

In the present paper we propose a new method for estimating the mass
of clusters. We intend to infer total cluster mass from the knowledge
of the kinematics in the outskirts, where the matter has not yet
reached equilibrium.
The key of our method is the analysis of filamentary and sheetlike
structures flowing outside the cluster.  We apply our method for the
total virial mass estimate to the Coma cluster, and we compare our
result with some of the previous ones in the literature. Our method
also provides an estimation of the orientation of the structures we find, in
the three dimensional space. 
This can be useful to identify a major merging plane, if a sufficient number of structures are detected and at least three of them are on the same plane.

The paper is organized as follows. In section 2 we derive the relation
between the velocity profile of galaxies in the outer region of
clusters and the virial cluster mass. In section 3 we propose a method
to detect filaments or sheets by looking at the observed velocity
field. In section 4 we test the method to a cosmological simulated
cluster-size halo and we present the result on the mass measurement. In section 5 we present the
structures we find around the Coma cluster and the Coma virial mass determination. 

\section{Mass estimate from the radial velocity profile}

Galaxy clusters are characterized by a virialized region where the
matter is approximately in dynamical equilibrium.
The radius that delimitates the equilibrated
cluster, i.e. the virial radius $r_{\rm v}$, is defined as the distance
from the centre of the cluster within which the mean density is
$\Delta$ times the critical density of the Universe $\rho_{c}$.
The virial mass is then given by
\begin{equation}
\label{eqn:vmass}
M_{\rm v}=\frac{4}{3}\pi\,r_{\rm v}^{3}\,\Delta\,\rho_{c} \, .
\end{equation}

The critical density is given by 
\begin{equation}
\label{eqn:vmass2}
\rho_{c}=\frac{3\,H^{2}}{8\pi\,G}\, ,
\end{equation}
where $H$ is the Hubble constant and $G$ the universal gravitational constant.

The circular velocity $V_{\rm v}$ at $r=r_{\rm v}$, i.e. the virial velocity, is defined as 
\begin{equation}
\label{eqn:vvel}
V_{\rm v}^{2}=\frac{G\,M_{\rm v}}{r_{\rm v}}.
\end{equation}

The immediate environments of galaxy clusters outside the virial radius are characterized by galaxies and groups of galaxies which are
falling towards the cluster centre. These galaxies are not part of the
virialized cluster, but they are gravitationally bound to it. The region where
the infall motion is most pronounced extends up to three or four times
the virial radius \citep{Mamon04,W05,Rines06,Cuesta08,Falco2013}.
At larger scales, typically beyond $6-10\,r_{\rm v}$, the radial motion
of galaxies with respect to the cluster centre, is essentially
dominated by the Hubble flow.
In the transition region between the
infall regime and the Hubble regime, the galaxies are flowing away
from the cluster, but they are still gravitationally affected by the
presence of its mass. At this scale, the gravitational
effect of the inner cluster mass is to perturb the simple Hubble
motion, leading to a deceleration.

The total mean radial velocity of galaxies outside clusters is therefore
the combination of two terms:
\begin{equation}
\label{eqn:vel}
\overline v_{\rm r} (r)=H\,r+\overline v_{\rm p} (r)\, ,
\end{equation}
the pure Hubble flow, and a mean negative
infall term $\overline v_{\rm p} (r)$, that accounts for the departure from
the Hubble relation. Section (\ref{infall}) is dedicated to the
characterization of the function $\overline v_{\rm p} (r)$.

The mean infall velocity depends on the halo mass, being more
significant for larger mass haloes.
Therefore, we can rewrite equation~(\ref{eqn:vel}) as
\begin{equation}
\label{eqn:velvir}
\overline v_{\rm r} (r,M_{\rm v})=H\,r+\overline v_{\rm p}(r,M_{\rm v})\, ,
\end{equation}
where we include the dependence on the virial mass $M_{\rm v}$.

Therefore, once we know the relation between $\overline v_{\rm p}$ and $M_{\rm v}$,
 equation~(\ref{eqn:velvir}) can be used to infer the virial mass of
 clusters.

In the next section, we will derive the equation that connects the
 peculiar velocity of galaxies $\overline v_{\rm p}$  with the virial mass of
 the cluster $M_{\rm v}$.

\subsection{Radial infall velocity profile}
\label{infall}
\begin{figure}
\centering
    \includegraphics[width=\hsize]{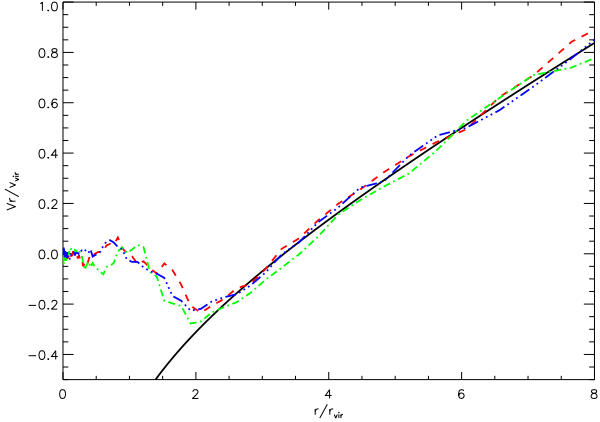}  
	\caption{Median radial
velocity profile for three samples of stacked simulated halos. The
virial masses for the samples are: $M_{\rm v}=0.8\times 10^{14}\,
M_{\odot}$ (blue, triple-dot dashed line),  $M_{\rm v}=1.1\times 10^{14}\,
M_{\odot}$ (green dot
dashed line), $M_{\rm v}=4.7\times 10^{14}\, M_{\odot}$ (red dashed
line). The black solid line is our simultaneous fit to the three profiles.}
\label{fig:3velocities}
\end{figure}

Simulations have shown a quite universal trend for the radial mean velocity
profile of cluster-size haloes, when normalized to their virial
velocities \citep{Prada06,Cuesta08}. This feature can be seen, for example, in Fig.~\ref{fig:3velocities}, where the median radial
velocity profile for three samples of stacked simulated haloes is
displayed. The units in the plot are the virial velocity $V_{\rm v}$
and virial radius $r_{\rm v}$.
The virial masses for the samples are: $M_{\rm v}=0.8\times 10^{14}\,
M_{\odot}$ (blue, triple-dot dashed line),  $M_{\rm v}=1.1\times 10^{14}\,
M_{\odot}$ (green dot
dashed line), $M_{\rm v}=4.7\times 10^{14}\, M_{\odot}$ (red dashed line). 
The cosmological N-body simulation we used is described in section~\ref{sec4}.

In order to derive an approximation for the mean velocity profile, the spherical collapse model has been
assumed in several works \citep{Pei06,Pei08,KN10,Nas11}.
Here we make a more conservative choice. We parametrize the infall
profile using only the
information that it must reach zero at large distances from the halo
centre, and then we fit the universal shape of the simulated haloes profiles. Therefore, we don't assume the spherical infall model.

In the region where the Hubble flow starts to dominate  and the
total mean radial velocity becomes positive, a good
approximation for the infall term is
\begin{equation}
\label{eqn:velpec}
\overline v_{\rm p} (r)\approx\,-v_0\,\left(\frac{r}{r_{\rm v}}\right)^{-b}\, ,
\end{equation}
with $v_0=a\,V_{\rm v}$, where $V_{\rm v}$ is the virial velocity, and $r_{\rm
v}$ is the virial radius.

We fit equation~(\ref{eqn:velpec}) to the three profiles in
Fig.~\ref{fig:3velocities} simultaneously, with $a$ and $b$ as free parameters. The fit is performed in the
range $r=3-8\,\rm r_{\rm v}$. The best fit is the black solid line,
corresponding to parameters: $a=0.8$ and $b=0.42$.

This allows to fix a universal shape for the mean velocity of the
infalling matter, as function of the virial velocity, i.e. the virial
mass, in the outer region of clusters.

\section{Filaments and sheets around galaxy clusters}

The method we propose for measuring the virial cluster mass, consists
in using only observed velocities and
distances of galaxies, which are outside the virialized part of the cluster, but whose
motion is still affected by the mass of the cluster.
Given the dependence of the infall velocity on the virial mass, we wish to estimate $M_{\rm v}$ by fitting the measured
 velocity of galaxies moving around the cluster with
 equations~(\ref{eqn:velvir}) and~(\ref{eqn:velpec}).

To this end, we need to select galaxies which are sitting, on average, in the transition
region of the mean radial velocity profile. For the fit to be
accurate, the galaxies should be spread over several megaparsec in radius. 

Observations give the two-dimensional map of clusters and their
surroundings, namely the projected radius of galaxies on the sky $R$,
and the component of the galaxy velocities along the line of sight
$v_{\rm los}$.
The reconstruction of the radial velocity profile would require 
the knowledge of the radial position of the galaxies, i.e. the
radius $r$.
The velocity profile that we infer from observations is also affected by the
projection effects.
If the galaxies were randomly located around clusters, the
projected velocities would be quite uniformly distributed, and we would not see any signature of the radial velocity
profile.
The problem is overcome because of the strong anisotropy of the matter
distribution. At several megaparsec away from the cluster centre, we
will select collections of galaxies bound into systems, as filaments or sheets.
The presence of such objects can break
the spatial degeneracy in the velocity space. 

In sections~(\ref{sec31}) and~(\ref{sec32}), we explain in details how such objects 
can be identified as filamentary
structures in the projected velocity space.

\subsection{Line of sight velocity profile}
\label{sec31}
In order to apply the universal velocity profile~(\ref{eqn:velpec}) to observations, we need to 	
transform the 3D radial equation~(\ref{eqn:vel}) in a 2D projected equation. 
We thus need to compute the line of sight velocity profile
$v_{\rm los}$  as function of the projected radius $R$.

\begin{figure}
\begin{center}
\centering
   \includegraphics[width=0.7\linewidth]{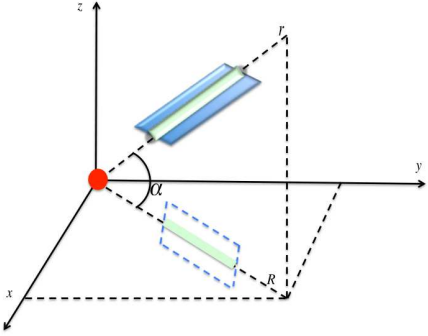}  
	\caption{Schematic drawing of a filament or a sheet
          in 3D with inclination $\alpha$
          between the radial distance $r$ and the projected radius
          $R$. The cluster is represented by the red circle in the
          centre of the frame. The $z$-axis corresponds to the observer
        line of sight.} 
\label{fig:drawing}
\end{center}
\end{figure}

Let's consider a filamentary structure forming an
angle $\alpha$ between the 3-dimensional radial position of galaxy members $r$ and the 2-dimensional projected radius $R$.
Alternatively, let's consider a sheet in the 3D space lying on a plan with inclination
$\alpha$ with respect to the plan of the sky (see the schematic Fig.~\ref{fig:drawing}).

The transformations between quantities in the physical space and in the
redshift space are

\begin{equation}
R=\cos\alpha\,r
\end{equation}
for the spatial coordinate, and

\begin{equation}
\label{eqn:vellos}
v_{\rm los} (R)=\sin\alpha\,v_{\rm r}(r)
\end{equation}
for the velocity.

By inserting equation~(\ref{eqn:velvir}) in equation~(\ref{eqn:vellos}), we
obtain the following expression for the line of sight velocity in the
general case:

\begin{eqnarray}
\label{eqn:vellos2}
v_{\rm los} (R,\alpha,M_{\rm v})=
\sin\alpha\,\left[H\,\frac{R}{\cos\alpha}+v_{\rm
 p}\left(\frac{R}{\cos\alpha},M_{\rm v}\right)\right] .
\end{eqnarray}

If we use our model for the infall term, given by
equation~(\ref{eqn:velpec}), the line of sight velocity profile in
equation~(\ref{eqn:vellos2}) becomes
\begin{eqnarray}
\label{eqn:vellos3}
v_{\rm los} (R,\alpha,M_{\rm v})
=\sin\alpha\,\left[H\,\frac{R}{\cos\alpha}-a\,V_{\rm v}\,\left(\frac{R}{\cos\alpha\,r_{\rm
        v}}\right)^{-b}\right] \, .
\end{eqnarray}

By using equation~(\ref{eqn:vellos3}),
it is, in principle, possible to measure both the virial
cluster mass $M_{\rm v}$ and the orientation angle $\alpha$ of the
structure. In fact, if we select a sample of galaxies which lie in a
sheet or a filament, we can fit their phase-space coordinates
($R,v_{\rm los}$) with equation~(\ref{eqn:vellos3}), where only two free
parameters ($\alpha,M_{\rm v}$) are involved. The identification of structures and
the accuracy on the mass estimate
require a quite dense sample of galaxies observed outside the cluster.

\subsection{Linear structures in the velocity field}
\label{sec32}
Our interest here is thus in finding groups of galaxies outside
clusters, that form a bound system with a relatively small dispersion in
velocity, and that lie on a preferential direction in the 3D space.
In particular, we are interested in such objects when
they are far enough from the cluster, to follow a nearly linear radial pattern in the
velocity space, corresponding to a decelerated Hubble flow.

We expect these objects to form filament-like structures in the
projected velocity space.
In fact, if we apply the formula in equation~(\ref{eqn:vellos2}) to galaxies with the same orientation angle
$\alpha$ within a small scatter, the radial velocity shape given by equation~(\ref{eqn:velvir}) is preserved.
Thus, these galaxies can be identified as they are collected on a line 
in the observed velocity space.

Nevertheless, we can look at the structure in the 2D map (the ($x,y$)
plane in Fig.~\ref{fig:drawing}). 
If all the selected galaxies lie on a line, within a small scatter, also in
the ($x,y$) plane, they can be defined as a filament. If they are
confined in a region within a small angular aperture, they might form a
sheet (see the Fig.~\ref{fig:drawing}). 
Complementary papers will analyze properties of such sheets \citep{Thejs,sparre2013,Wadekar}.

We want to point out here that Fig.~\ref{fig:drawing} describes
the ideal configuration for filaments and sheets to have a quasi-linear shape in the observed velocity
plane. Therefore, not all the filaments and sheets will satisfy this
requirement, i. e. not all the structures outside clusters can be
detected by looking at the velocity field. 

Our method for identifying these objects
is optimized towards structures which are narrow in velocity space,
while still containing many galaxies, and therefore which are closer to face-on than edge-on.
It consists in selecting a region in
the sky, and looking for a possible presence of an
overdensity in the
corresponding velocity space. We will describe the method in details
in the next section.

\section{Testing the method on Cosmological Simulation} 
\label{sec4}
As a first test of our method, we apply it to a cluster-size halo from
a cosmological N-body simulation of pure Dark Matter (DM).

The N-body simulation is based on the $WMAP3$ cosmology. The
cosmological parameters are $\Omega_{\rm M}=0.24$ and
$\Omega_{\Lambda}=0.76$, and the reduced Hubble parameter is $h=0.73$. The
particles are confined in a box of size $160\,h^{-1}$ Mpc. The particle
mass is $3.5\times\,10^{8}\,\rm  M_\odot$, thus there are $1024^{3}$ particles in
the box. The evolution is followed from the initial redshift $z=30$,
using the MPI version of the ART code \citep{Kra1997,G2008}.  The algorithm used to
identify clusters is the hierarchical friends-of-friends (FOF) with a
linking length of 0.17 times the mean interparticle distance. The
cluster centres correspond to the positions of the most massive
substructures found at the linking length eight times shorter than the mean interparticle distance.
We define the virial radius of
halos as the radius containing an overdensity of $\Delta=93.8$ relative
to the critical density of the Universe. More details on the
simulation can be found in \citep{Wojtak08}.

For our study, we select, at redshift
$z=0$, a halo of virial quantities $M_{\rm v}=4.75\times\,10^{14}\, \rm  M_\odot$, $\rm r_{\rm v}=2.0\,Mpc$ and
$V_{\rm v}=1007.3\, \rm km/s$. 

We treat the DM particles in the halo as galaxies from observations.                           
The first step is to project the 3D halo as we would see it on the
sky. We consider three directions as possible lines of sight. For
each projection, we
include in our analysis all galaxies in the box $x=[-20,20]\,\rm Mpc$ and
$y=[-20,20]\,\rm Mpc$, where $x,y$ are the two directions perpendicular to
the line of sight. 

The method described in the next section is applied to all the three projections.

\subsection{Identification of filaments and sheets from the velocity
  field} 

\begin{figure}
\centering
    \includegraphics[width=\hsize]{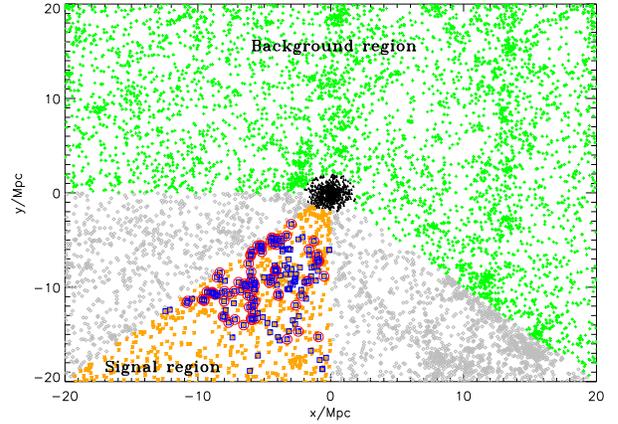}  
	\caption{Two-dimensional projection of the simulation box,
          centered on the selected
         simulated halo. The black triangles represent the particles inside
      the virial radius of the halo. The orange points belong to one
      of the eight wedges we select in the $(x,y)$ plane. The background for the
        selected wedge is given by the green crosses. The two wedges
        adjacent to the selected wedge, gray diamonds, are
        excluded from the analysis. In the selected wedge, we identify
      a sheet that is represented by the red circles. The blue squares
      correspond to the total overdensity we find in the wedge, with
      the method described in the text.}
\label{fig:haloxy}
\end{figure}

\begin{figure}
\centering
    \includegraphics[width=\hsize]{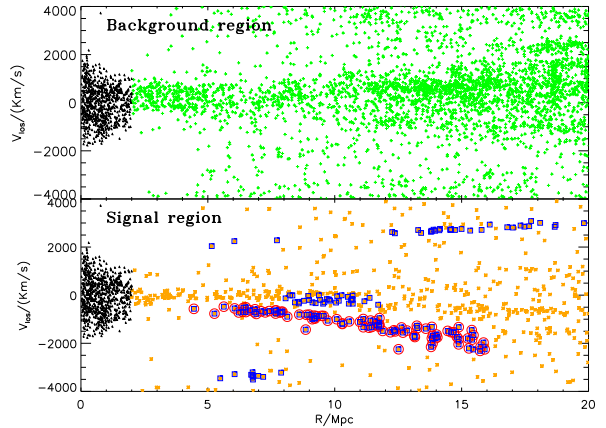}  
\caption{Line of sight velocity $v_{\rm los}$ as function of the
  projected distance $R$ from the centre of the simulated halo. \emph{Upper panel}: The background in the
          analysis is represented by the green crosses. The black
          triangles are all the particles within the virial
          radius. \emph{Bottom panel}: The orange points represent
          our signal, i.e. the selected wedge. The blue points correspond
        to the overdensity in the wedge. The only almost straight
        inclined line is shown in red circles. We identify this
        filamentary-like structure as a sheet.}
\label{fig:fil1}
\end{figure}

Our goal is to find structures confined in a relatively small area
in the $(x,y)$ plane. To this end, we split the spatial distribution
into eight two-dimensional wedges (for example in Figure~\ref{fig:haloxy}
the orange points represent one of the wedges) and
we look at each of
them in the  $(R,v_{\rm los})$-space (for example in Fig.~\ref{fig:fil1} we look at the
orange wedge in Fig.~\ref{fig:haloxy}, in the velocity space), where we aim to look for overdensities. 

We confine the velocity field to the box: $v_{\rm los}=[-4000,4000]\,\rm km/s$  and  $\rm R=[4,20]\,\rm Mpc$, and we divide the box into $50$ cells, $4\, \rm Mpc$ large and $400\, \rm km/s$
high.

For each of the selected wedges, we want to compare the  galaxy number density $n_{i}$ in each cell $i$,
with the same quantity calculated for the the rest of the wedges in
the same cell. 
More
precisely, in each cell, we calculate the mean of the galaxy number
density of all the wedges but the selected one.
This quantity acts as background for the
selected wedge, and we refer to it as $n^{bg}_{i}$.

In Fig.~\ref{fig:haloxy}, the wedge under analysis
is represented by the orange points, and the background by the green
points. We exclude from the background the two wedges adjacent to the
selected one (gray points in Fig.~\ref{fig:haloxy}).
We need this step because, if any structure is sitting in the selected
wedge, it might stretch to the closest wedges. 

The overdensity in the cell $i$ is evaluated as 
\begin{equation}
\label{eqn:od}
m_{i}= \frac{n_{i}-n^{bg}_{i}}{n^{bg}_{i}} \, ,
\end{equation}
and we calculate the probability density $p(m_{i})$
for the given wedge. 
We take only the cells in the top $1\,\sigma$ region of the
probability density distribution, i.e. where the integrated probability is above
$(100-16.8)\%$, in order to reduce the background noise.
Among the galaxies belonging to the selected cells, we take the ones
lying on inclined lines within a small scatter, while we remove the unwanted
few groups which appear as blobs or as horizontal strips in the
$(R,v_{\rm los})$-space. 
We apply this selection criterion because we are interested in extended
structures which have a coherent flow relative to the cluster.

This method leaves us with only one structure inside the wedge in
Fig.~\ref{fig:haloxy} (red points). 
It is a sheet, as it
appears as a two-dimensional object on the sky,
opposed to a filament which should appear one-dimensional.
We see such sheet only
in one of the three projections we analyse.
The bottom panel of Fig.~\ref{fig:fil1} shows the 
velocity-distance plot corresponding to all the
galaxies belonging to the selected wedge (orange points), while the selected strips
of galaxies are shown as blue points. The desired sheet (red points) is an almost straight inclined
line crossing zero velocity roughly near 5-10 Mpc and contains 88
particles. The background wedges are
displayed in the upper panel of Fig.~\ref{fig:fil1}.

\subsection{Analysis and result} 
\label{sec42}
Having identified one sheet around the simulated halo, we can now extract the halo mass, using
the standard Monte Carlo fitting methods.
We apply the Monte Carlo Markov chain to the galaxies belonging to the
sheet. The
model is given by equation~(\ref{eqn:vellos3}), where the free parameters are
$(\alpha,\rm M_{\rm vir})$. We set $\Delta=93.8$ and $H=73\,\rm
km/(s\,Mpc)$, as these are the values set in the cosmological simulation.
We run one chain of $5000$ combinations of parameters and then we
remove the burn-in points.
\begin{figure}
\begin{center}
\includegraphics[width=1.1\linewidth]{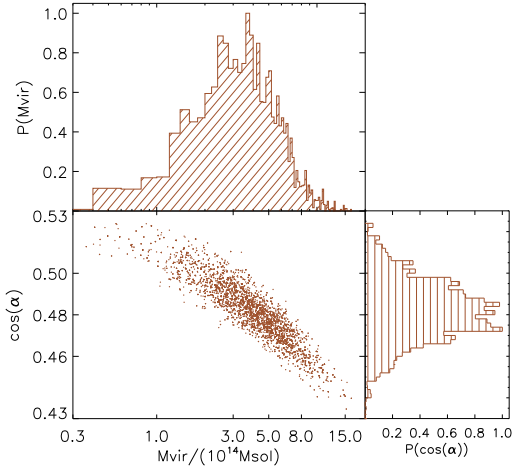}
\end{center}
	\caption{Result of the Monte Carlo Markov chain applied to the
          sheet found outside the simulated halo. \emph{Central panel}: Scatter plot of the two free parameters
         (${\rm cos}(\alpha),M_{\rm vir}$) obtained by the MCMC
         chain. \emph{Upper panel}: Probability density function of
         the virial mass. \emph{Left panel}: Probability density
         function of the viewing angle. The initial number of points is
         5000 and we remove the points of burn-in. The mean value for
         the virial mass and the cosine of the angle are $M_{\rm
  vir}=(4.3\pm\,2.2)\times\,10^{14}\,\rm  M_\odot$ and ${\rm cos}(\alpha)=0.48\pm\,0.02$, which are
comparable to the true halo virial mass $M_{\rm
  vir}=4.75\times\,10^{14}\,\rm  M_\odot$ and angle ${\rm cos}(\alpha)=0.5$.} 
\label{fig:scatterplotSIM}
\end{figure}

\begin{figure}
\begin{center}
    \centering
    \includegraphics[width=0.5\textwidth]{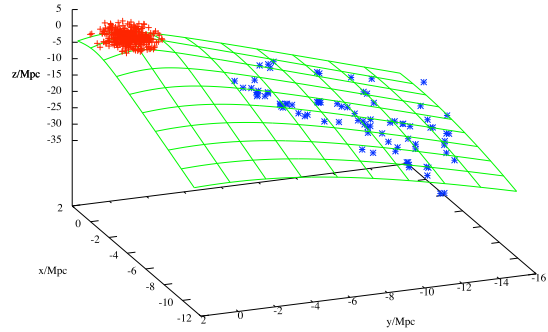}
	\caption{The sheet we found outside the simulated halo in the three-dimensional space. The $z$-axis corresponds to the line of
          sight direction.
The blue points represent the particles
          belonging to the sheet, and the green plane is the best fit
          for the sheet's plane, corresponding to
          $\alpha=-1.05$ (${\rm
  cos}(\alpha)=0.5$) rad. The red points represent the particles
          within the virial radius of the halo.} 
\label{fig:sheetSIM}
\end{center}
\end{figure}

In Fig.~\ref{fig:scatterplotSIM} we show the scatter plot on the
plane of the two
parameters, and the one-dimensional probability
distribution functions of the virial mass and the orientation
angle. The mean value for the virial mass is $M_{\rm
  vir}=(4.3\pm\,2.2)\times\,10^{14}\,\rm  M_\odot$, which is
comparable to the true halo virial mass $M_{\rm vir}=4.75\times\,10^{14}\,\rm  M_\odot$.
The mean value for the cosine of the angle between $R$ and $r$ is
${\rm cos}(\alpha)=0.48\pm\,0.02$, corresponding to 
$\alpha=-1.07\pm\,0.02$~rad. 
In Fig.~\ref{fig:sheetSIM} we show the sheet in the 3D space (blue points). The
best fit for the plane where the sheet is 	
laying, is shown as the green plane, and the corresponding angle is
$\alpha=-1.05$~rad, giving ${\rm cos}(\alpha)=0.5$.
Our estimation is thus consistent, within the statistical error, with the true orientation of the sheet
in 3D.

Although our method provides the correct halo mass and orientation angle within the
errors, the results slightly underestimate the true values, for both parameters.
Systematic errors on the mass and angle estimation might be due to the non ideal
shape of the structures. The sheet we find has finite thickness, and it
is not perfectly straight in the 3D space. The closer the
detected structure is to an ideal infinite thin and perfectly
straight object, the smaller the errors would be. 
Another problem might reside in the assumption of spherical
symmetry. The median radial velocity profile of a stack of haloes, might slightly differ
from the real velocity profile of each individual halo.
Intrinsic scatter of the simulated infall velocity profiles leads 
to additional systematic errors on the determination of the best fitting parameters. 
Our estimate of this inaccuracy yields $50\%$ for the virial mass and 
 $2.5\%$ for the angle.

\begin{figure}
\centering
\begin{minipage}[b]{.5\textwidth}
\includegraphics[width=0.8\linewidth]{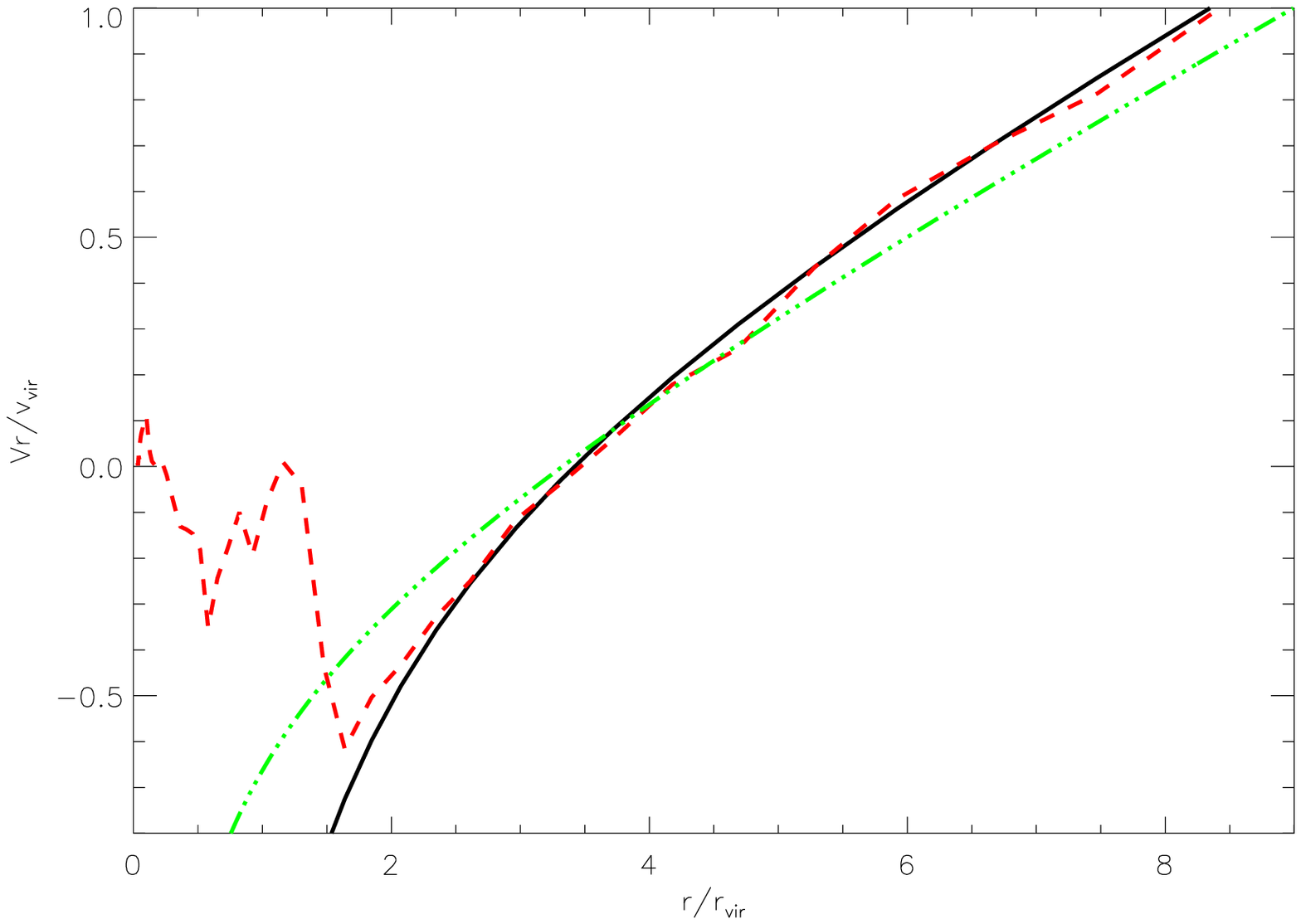}
\end{minipage}
\quad
\begin{minipage}[b]{.5\textwidth}
\includegraphics[width=1.0\linewidth]{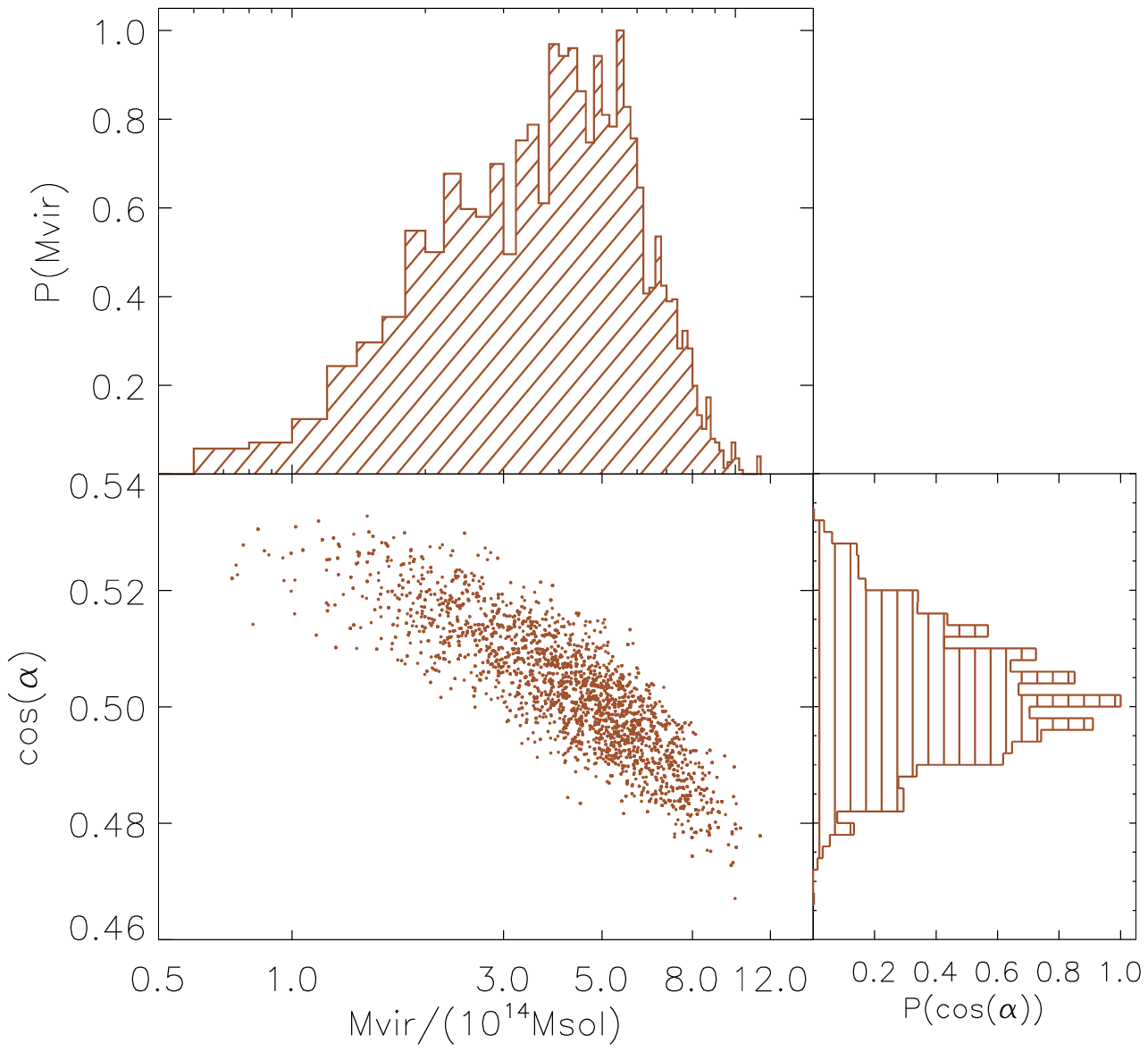}
\caption{
The top figure shows the median radial
velocity profile for the simulated halo (red dashed line). The black
solid line is our fit to the profile. The green dot-dashed line is the
universal radial profile showed in Fig.~\ref{fig:3velocities}.
The bottom figure shows the result of the Monte Carlo Markov chain applied on the
          sheet found around the simulated halo, using the fit to the mean
          velocity profile of the halo (top figure). \emph{Central panel}: Scatter plot of the two free parameters
          (${\rm cos}(\alpha),M_{\rm vir}$) obtained by the MCMC
         chain. \emph{Upper panel}: Probability density function of
         the virial mass. \emph{Left panel}: Probability density
         function of the viewing angle. The initial number of points is
         5000 and we remove the points of burn-in. The mean value for
         the virial mass is $M_{\rm
  vir}=(4.67\pm\,1.9)\times\,10^{14}\,\rm  M_\odot$, which is very close
to the true halo virial mass $M_{\rm vir}=4.75\times\,10^{14}\,\rm
M_\odot$. 
The mean value for the cosine of the angle is
 ${\rm cos}(\alpha)=0.5\pm\,0.01$, in agreement with the real
 value ${\rm cos}(\alpha)=0.5$.}
\label{fig:ScatterplotHALOreal}
\end{minipage}
\end{figure}

The presence of this systematic is confirmed by
Fig.~\ref{fig:ScatterplotHALOreal}. The bottom panel represents our
result of the sheet analysis, when using a fit to the real mean radial velocity of the halo, which is
shown in the upper panel. 
The best fit parameters to the radial
velocity profile of the halo, with equation~(\ref{eqn:velpec}),
are $a=1.5$ and $b=0.89$. 
In Fig.~\ref{fig:ScatterplotHALOreal}, the
black solid line is the fit to the halo velocity profile (red dashed
line) and the green
dot-dashed line is the universal velocity profile used in the previous
analysis.  
The two profiles overlap in the range $\approx\,3-5\,r_{\rm v}$, but they
slightly differ for larger distances, where our sheet is actually sitting.
Replacing the universal radial velocity profile with the true
one, eliminates the 	
small off set caused by the departure of the two profiles.
In the new analysis, the mean value for the virial mass is $M_{\rm
  vir}=(4.67\pm\,1.9)\times\,10^{14}\,\rm  M_\odot$, while
the mean value for the cosine of the angle between $R$ and $r$ is
${\rm cos}(\alpha)=0.5\pm\,0.01$.
They are in very good agreement with the true values of the
parameters  
$M_{\rm vir}=4.7\times\,10^{14}\,\rm  M_\odot$ and ${\rm
  cos}(\alpha)=0.5$.

\section{Result on Coma Cluster}
In this section, we will apply our method to real data of the Coma
cluster.

We search for data in and around the Coma Cluster in the SDSS database \citep{aa2009}.
We take the galaxy NGC 4874 as the centre of the Coma cluster \citep{kent82}, which has coordinates: 
RA=12h59m35.7s, Dec=+27deg57'33''.
We select galaxies within 18 degrees from the position of the Coma centre and with velocities between 3000 and 11000 km/s. 
The sample contains 9000 galaxies.

We apply the method for the identification of structures outside
clusters to the Coma data. We detect two galactic sheets in the environment of Coma. We denote our
sheets as \emph{sheet 1} and \emph{sheet 2}.

\begin{figure}
\centering
\begin{minipage}[b]{.5\textwidth}
\includegraphics[width=1.0\linewidth]{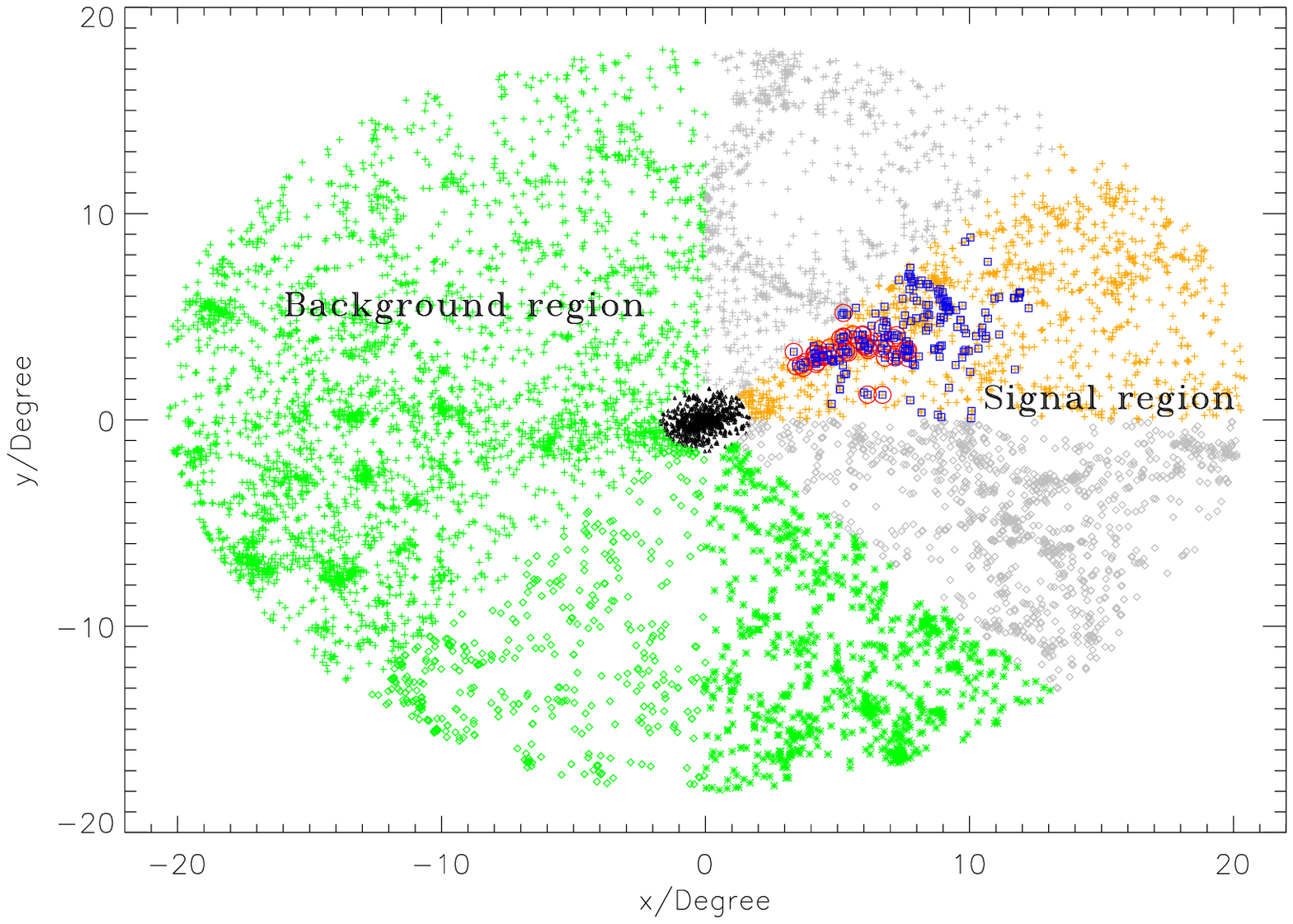}
\end{minipage}
\quad
\begin{minipage}[b]{.5\textwidth}
\includegraphics[width=1.0\linewidth]{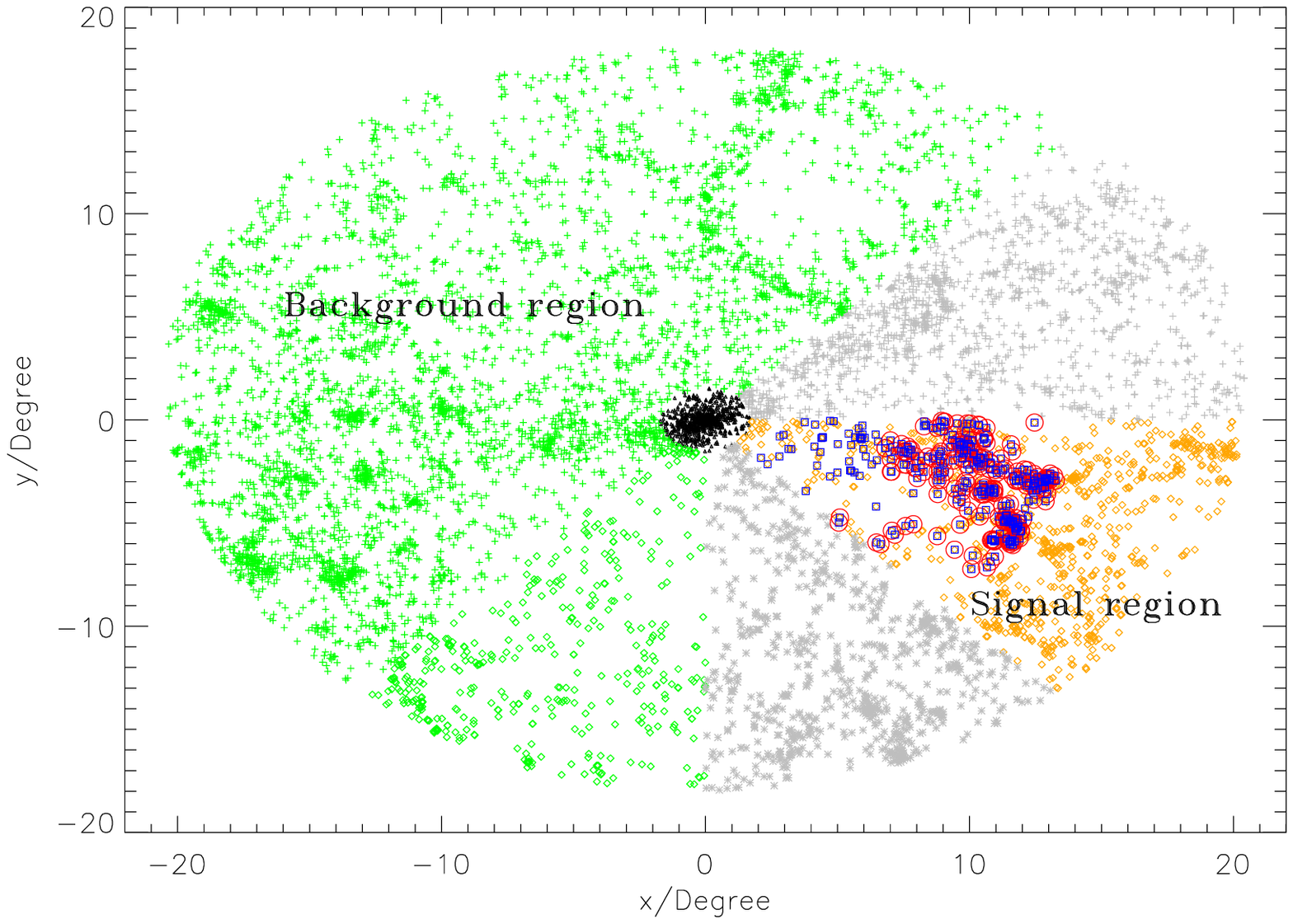}
\caption{Sky map of the Coma cluster. The top figure shows the
  \emph{sheet 1} and the bottom figure shows the \emph{sheet 2}. The black triangles represent the particles inside
      the virial radius of Coma. The orange points belong to one
      of the eight wedges we select in the $(x,y)$ plane. The background for the
        selected wedge is given by the green crosses. The two wedges
        adjacent to the selected wedge, gray diamonds, are
        excluded from the analysis. In the selected wedge, we identify
      a sheet that is represented by the red circles. The blue squares
      correspond to the total overdensity we find in the wedge, with
      the method described in the text.}
\label{fig:Comaxy}
\end{minipage}
\end{figure}

\begin{figure}
\centering
\begin{minipage}[b]{.5\textwidth}
\includegraphics[width=1.0\linewidth]{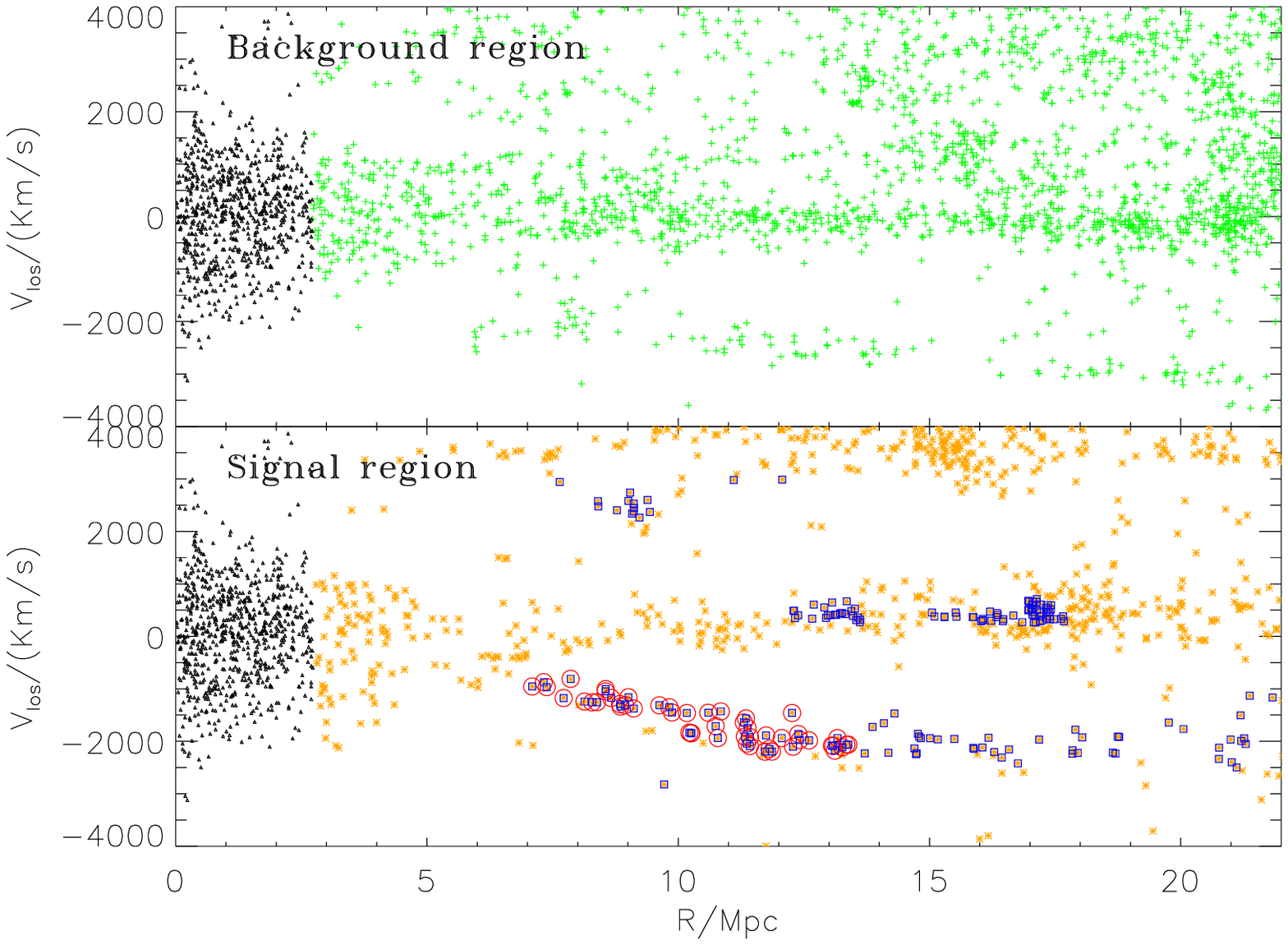}
\end{minipage}
\quad
\begin{minipage}[b]{.5\textwidth}
\includegraphics[width=1.0\linewidth]{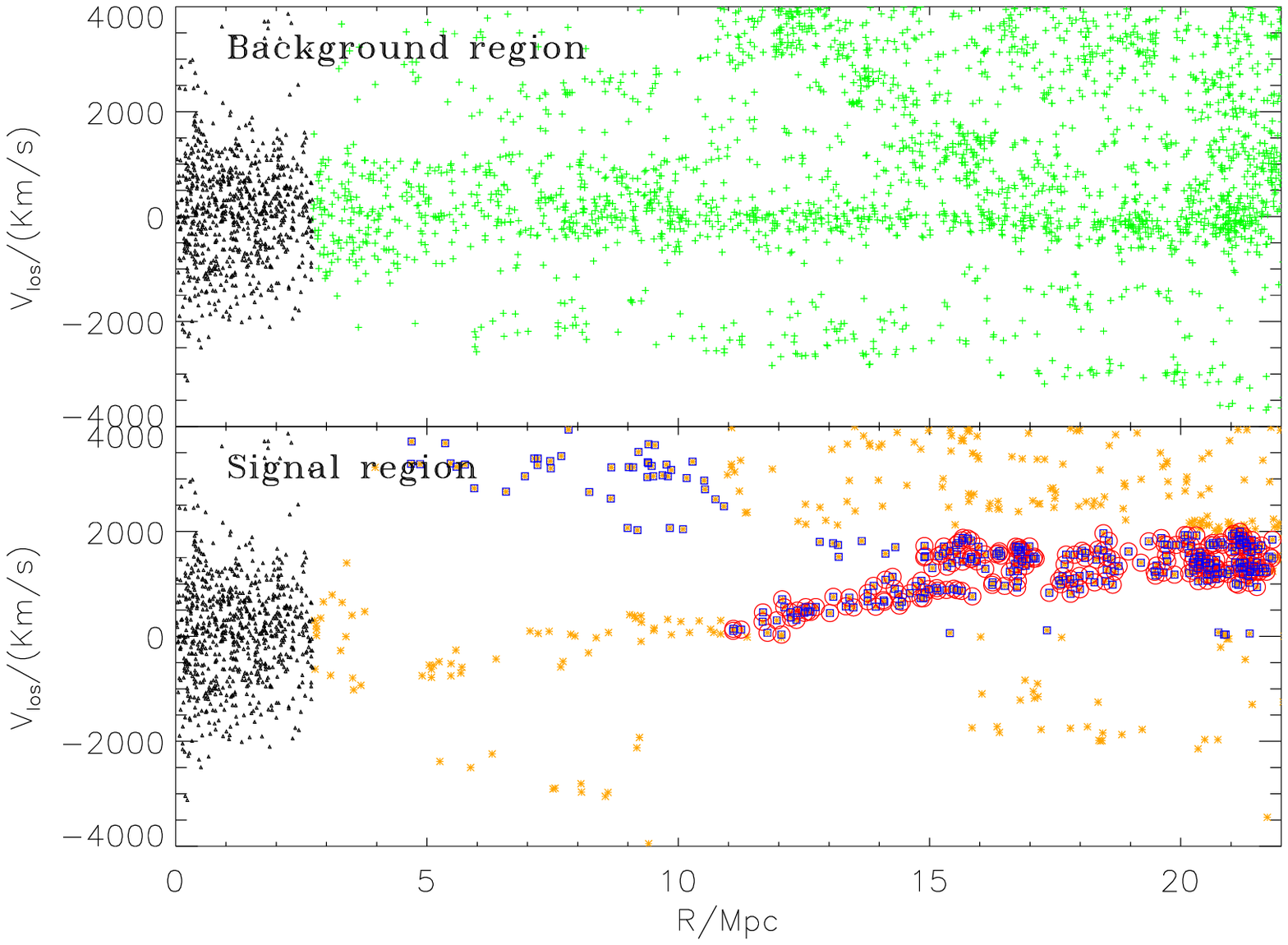}
\caption{Line of sight velocity $v_{\rm los}$ as function of the
  projected distance $R$ from the centre of Coma. The velocities are
  scaled by the velocity of Coma $v_{\rm Coma}=4000$km/s.  The top figure shows the
  \emph{sheet 1} and the bottom figure shows the \emph{sheet 2}.
 \emph{Upper panels}: The background in the
          analysis is represented by the green crosses. The black
          triangles are all the galaxies within $r=2.5$ Mpc. \emph{Bottom panel}: The orange points represent
          the signal, i.e. the selected wedge. The blue points correspond
        to the overdensity. The almost straight
        inclined lines are shown in red circles. We identify these
        filamentary-like structures as sheets.}
\label{fig:Comav}
\end{minipage}
\end{figure}

Fig.~\ref{fig:Comaxy} shows the Coma cluster and its environment up to 18
degrees from the cluster centre. The number
of galaxies with spectroscopically measured redshifts within $2.5\,$Mpc, which is roughly the virial radius of Coma, is 748. These
galaxies are indicated as black triangles. The sheets are the red
circles. The upper panel refers to
the \emph{sheet 1}, which contains
51 galaxies. The bottom panel refers to the \emph{sheet 2}, which is
more extended and contains 228 galaxies. 
In Fig.~\ref{fig:Comav}, we show the sheets in the velocity
space. They both appear as inclined straight lines. The \emph{sheet 1}
goes from $\approx\,7\,$Mpc to  $\approx\,14\,$Mpc. As the velocities
are negative, the sheet is between us and Coma. The \emph{sheet 2}
goes from $\approx\,11\,$Mpc to  $\approx\,22\,$Mpc. As the velocities
are positive, the sheet is beyond Coma.

As we did for the cosmological simulation, we have removed the
collections of galaxies which are horizontal groups in ($R,v_{\rm
  los}$)-space by hand. For example, in the case of the \emph{sheet 1} in the upper
panel of Fig.~\ref{fig:Comav}, we define the sheet only by including 
the inclined pattern and therefore, by excluding the horizontal part of
the strip.

\begin{figure}
\centering
\begin{minipage}[b]{.5\textwidth}
\includegraphics[width=1.1\linewidth]{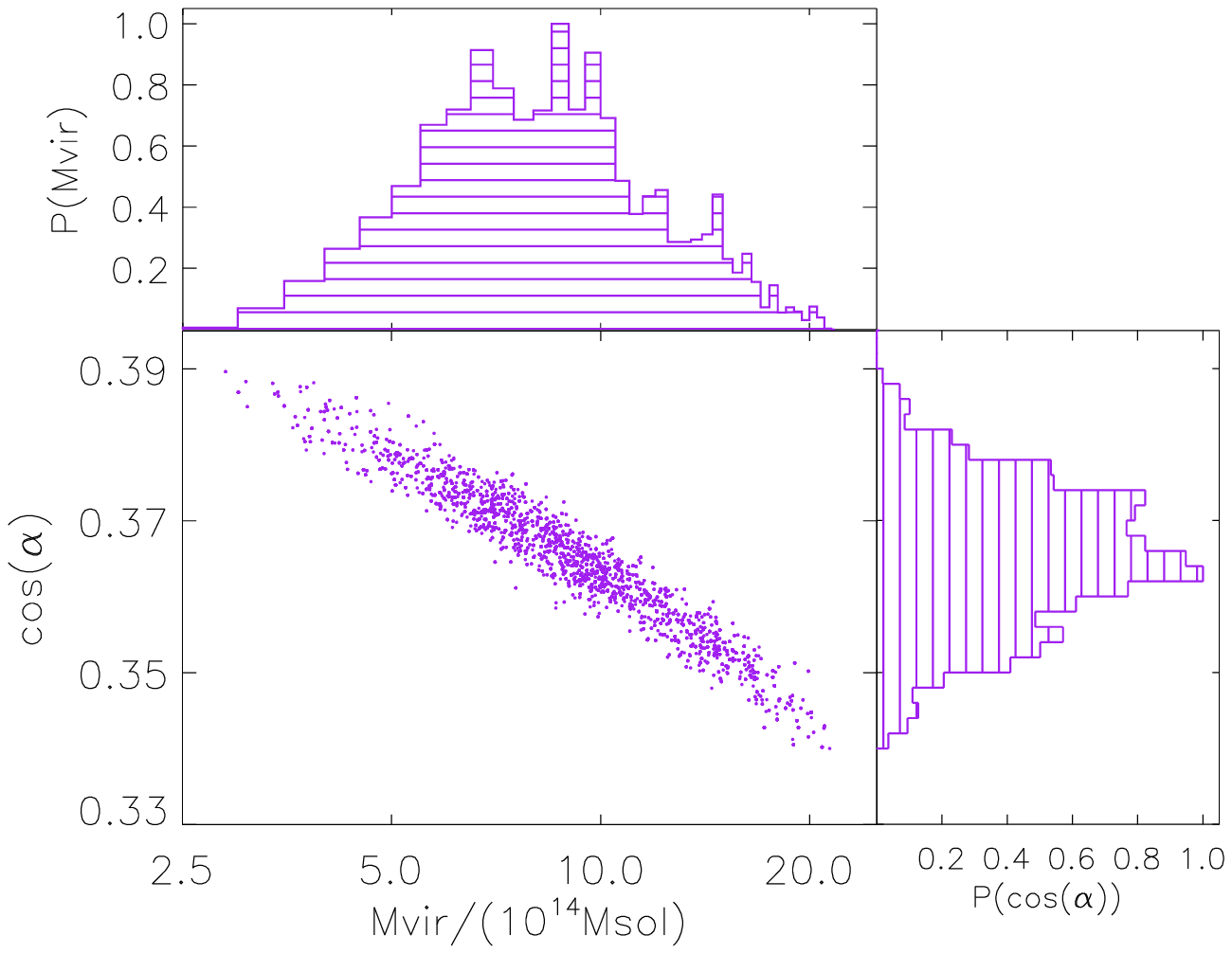}
\end{minipage}
\quad
\begin{minipage}[b]{.5\textwidth}
\includegraphics[width=1.1\linewidth]{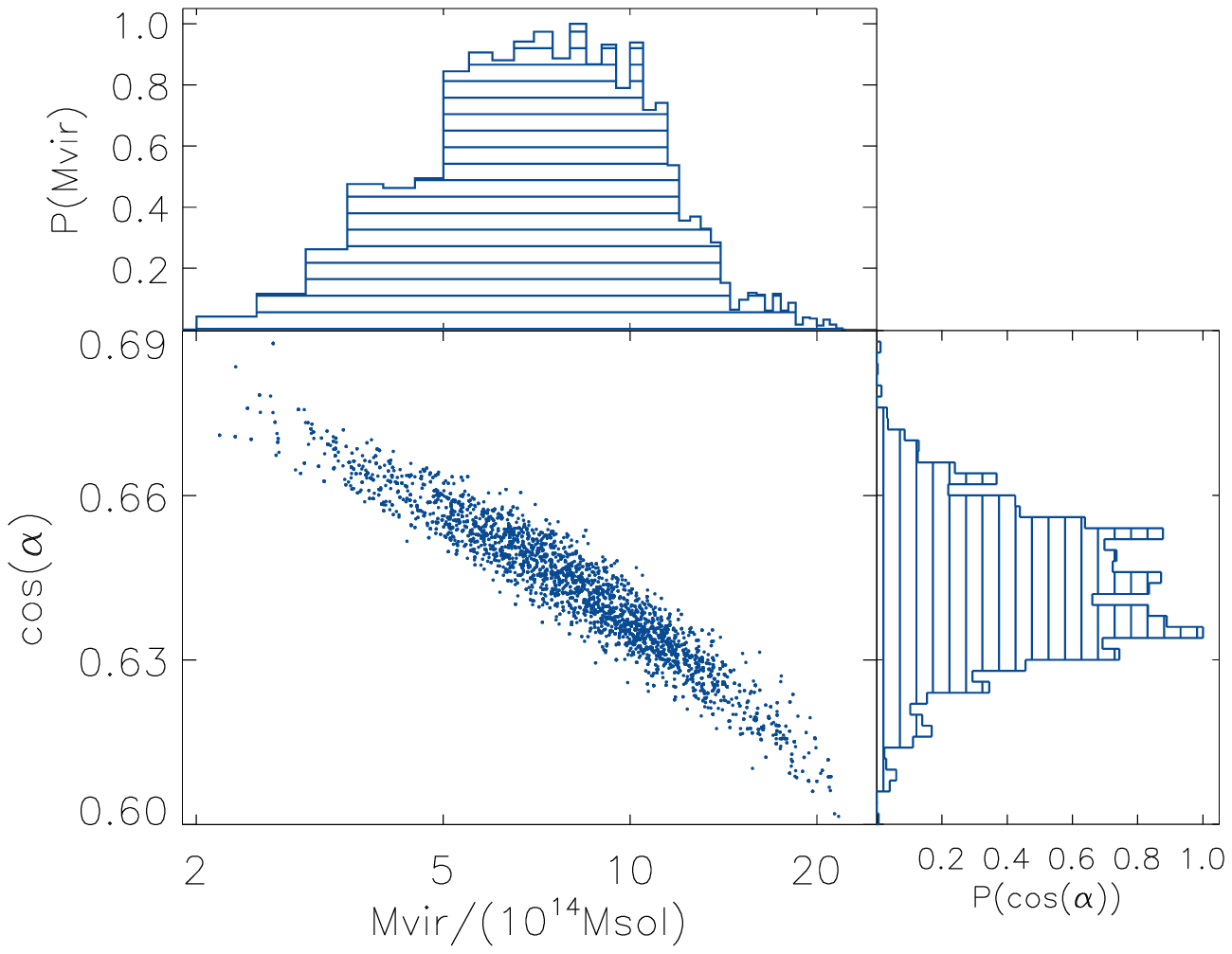}
\caption{Result of the Monte Carlo Markov chain applied to the two sheets found outside the Coma cluster.  The top figure refers to the
  \emph{sheet 1} and the bottom figure refers to the \emph{sheet
    2}. \emph{Central panels}: Scatter plot of the two free parameters
  (${\rm cos}(\alpha),M_{\rm vir}$) obtained by the MCMC
         chain. \emph{Upper panels}: Probability density function of
         the virial mass. \emph{Right panels}: Probability density
         function of the viewing angle. The initial number of points is
         5000 and we remove the points of burn-in.}
\label{fig:ScatterplotCOMA}
\end{minipage}
\end{figure}

We then fit the line of sight velocity profiles of the two sheets with
equation~(\ref{eqn:vellos3}).  We set $\Delta=93.8$ and $H=73\,\rm
km/(s\,Mpc)$, as for the cosmological simulation.

In Fig.~\ref{fig:ScatterplotCOMA} we show the scatter plot on the
plane of the two
parameters $({\rm cos}(\alpha),\rm M_{\rm vir})$, and the one-dimensional probability
distribution functions of the virial mass and the orientation
angle, for both the sheets. The angle $\alpha$ can be very
different for different sheets, as it only depends on the position of
the structure in 3D. Instead, we expect the result on the cluster mass $M_{\rm
  vir}$ to be identical, as it refers to the same cluster.

\begin{figure}
\centering
    \includegraphics[width=\hsize]{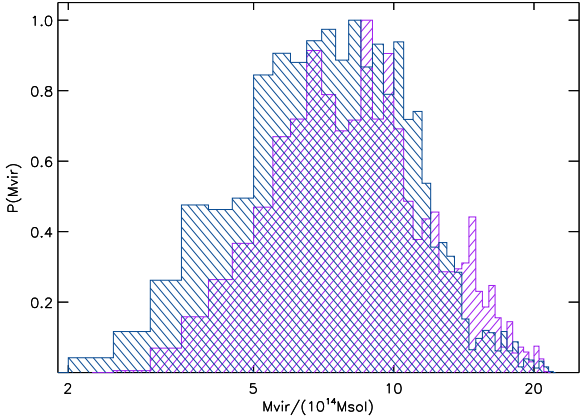}  
	\caption{The probability density function of the Coma virial
          mass, derived through the sheet technique. 
The distribution coming from the \emph{sheet 2} is the blue one,
slightly to the left. The violet slightly narrower distribution
corresponds to the \emph{sheet 1}.
The best mass estimate based on
  these measurement is: $M_{\rm vir}=(9.2\pm\,2.4)\times\,10^{14}\rm
  M_\odot$.}
\label{fig:masscoma}
\vspace{-0.2in}
\end{figure}

In Fig.~\ref{fig:masscoma}, we overplot the probability
distributions for the virial mass of Coma, from the analysis of the two
sheets. The two probability distributions are very similar. 
The mean value of the virial mass is $M_{\rm vir}=(9.7\pm\,3.6)\times\,10^{14}\rm  M_\odot$ for the \emph{sheet 1} and
$M_{\rm vir}=(8.7\pm\,3.3)\times\,10^{14}\rm  M_\odot$ for the
\emph{sheet 2}. When applying equation~(\ref{eqn:vmass}), these values
give a virial radius of $r_{\rm
  vir}=2.5\,$Mpc and $r_{\rm
  vir}=2.4\,$Mpc, respectively. The best mass estimate based on the
combination of
  these measurements is: $M_{\rm vir}=(9.2\pm\,2.4)\times\,10^{14}\rm
  M_\odot$.

Our result is in good agreement with previous estimates of the
Coma cluster mass. 
In \cite{Hug89}, they obtain a virial mass
$M_{\rm vir}=(13\pm\,2)\times10^{14}\,M_{\odot}$ from their X-ray
study. From the galaxy kinematic analysis, \cite{LM03} report a virial
mass $M_{100}=(15\pm\,4.5)\times10^{14}\,M_{\odot}$, corresponding to
a density contrast $\Delta=100$, which is very close to our value. \cite{Geller99} find a mass $M_{200}=15\,\times\,10^{14}\,M_{\odot}$,
corresponding to a density contrast $\Delta=200$. The weak
lensing mass estimate in \cite{kubo07} gives $M_{200}=2.7^{+3.6}_{-1.9}\,\times\,10^{15}\,M_{\odot}$.

The mean value for cosine of the orientation angle is
${\rm cos}(\alpha)=0.36\pm\,0.01$, corresponding to 
$\alpha=-1.2\pm0.01$~rad, for the \emph{sheet 1} and
${\rm cos}(\alpha)=0.64\pm\,0.02$, corresponding to
$\alpha=0.87\pm\,0.02$~rad, for the \emph{sheet 2}.   
These results are affected by a statistical error of $50\%$ for the
mass and $2.5\%$ for the angle, as discussed
in Section~\ref{sec42}.

\begin{figure}
\begin{center}
    \includegraphics[width=0.5\textwidth]{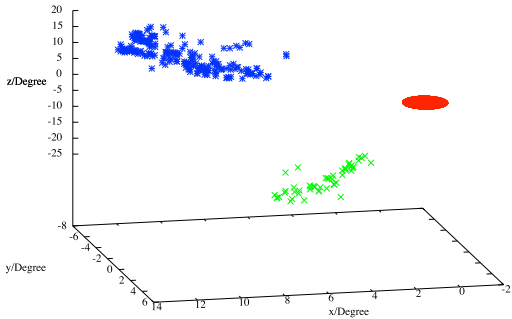}
	\caption{The sheets we found outside the Coma cluster in the three-dimensional space. The blue and the green points represent the particles
          belonging to the \emph{sheet 1} and the \emph{sheet 2},
          respectively. The Coma cluster is indicated as a red sphere centered
          at ($x,y,z$)=(0,0,0).} 
\label{fig:COMAsheets}
\end{center}
\end{figure}

The value obtained for the orientation $\alpha$ of a sheet corresponds to the
mean angle of all the galaxies belonging to the sheet.
By knowing $\alpha$, we can calculate the
corresponding coordinate along the line of sight for all the galaxies, and
therefore, we reconstruct the three dimensional map of the two
structures, as shown in Fig.~\ref{fig:COMAsheets}. The sheets we
find are lying on two different planes.

\section{Summary and Conclusion}
The main purpose of this paper is to propose and test a new method for
the mass estimation of clusters within the virial radius. The
idea is to infer it only from the kinematical data of structures
in the cluster outskirts.
 
In the hierarchical
scenario of structure formation, galaxy clusters are located at the intersection of filaments and sheets.
The motion of such non-virialized structures is thus affected by the
presence of the nearest massive cluster.

We found that modeling the kinematic data of these objects leads to an estimation of the neighbor
cluster mass. The gravitational effect of the cluster mass is to perturb the pure Hubble motion, leading
to a deceleration. Therefore, the measured departure from the Hubble
flow of those structures allows us to infer the virial
mass of the cluster.
We have developed a technique to detect the presence of structures outside galaxy clusters, by looking
at the velocity space. We underline that the proposed technique doesn't aim
to map all the objects around clusters, but it is limited to finding
those structures that are suitable for the virial cluster mass
estimation. 

Our mass estimation method doesn't require
the dynamical analysis of the virialized region of the cluster,
therefore it is not based on the dynamical equilibrium
hypothesis. However, our method rely on the assumption of
spherical symmetry of the system. In fact, we assume a radial
velocity profile. Moreover, our method is biased by fixing the 
phenomenological fit  to the radial infall velocity profile of
simulation, as universal infall profile.
From the practical point of view, this technique requires gathering
galaxy positions and velocities in the outskirts of galaxy clusters,
very far away from the cluster centre. A
quite dense sample of redshifts is needed, in order to identify the possible
presence of structures over the background. 
Once the structures are detected, the fit to their line of sight velocity
profiles has to be performed. The fitting procedure involves only two free
parameters: the virial mass of the cluster and the orientation angle
of the structure in 3D. This makes the estimation of the
virial cluster mass quite easy to obtain. 

We have analysed cosmological simulations first, in order to test both the
technique to identify structures outside clusters and the method to
extract the cluster mass. We find one sheet outside the selected
simulated halo, and we infer the correct halo mass and sheet
orientation angle, within the errors.

We then applied our method to the Coma cluster.
We have analysed the SDSS data of projected distances and velocities,
up to $20\,$Mpc far from the Coma centre.
Our work led to the detection of two galactic sheets in the environment of the
Coma cluster. The estimation of the Coma cluster mass through the
analysis of the two sheets, gives $M_{\rm
  vir}=(9.2\pm\,2.4)\times\,10^{14}\rm  M_\odot$. This value is
in agreement with previous results from the standard methods. We note
however that our method tends to underestimate the Coma virial mass,
compared with previous measurements, which either assume equilibrium
or sphericity.

In the near future, we aim to apply our technique to other surveys,
where redshifts at very large distances from the clusters centre are
available.
If a large number of sheets and filaments will be found, 
our method could also represent a tool to deproject the  spatial
distribution of galaxies outside galaxy clusters into the
three-dimensional space.

\vspace{-0.05in}

\section{Acknowledgements}

The authors thank Stefan Gottloeber, who kindly agreed for one of the
CLUES simulations (http://www.clues-project.org/simulations.html) to be used in the paper. 
The simulation has been performed at the Leibniz Rechenzentrum (LRZ), Munich.
The Dark Cosmology Centre is funded by the Danish National Research Foundation.

\bibliography{filaments1}

\end{document}